
%
%
%
%
%
%
%
\def\standardrisposta{s }\def\reducedrisposta{r }
\def\doublerisposta{d }\def\cartarisposta{e }\def\amsrisposta{y }
\newcount\ingrandimento \newcount\sinnota \newcount\dimnota
\newcount\unoduecol \newdimen\collhsize \newdimen\tothsize
\newdimen\fullhsize \newcount\controllorisposta
\newskip\infralinea  \global\controllorisposta=0
\message{ ********    Welcome to PANDA macros (Plain TeX, AP, 1991)}
\message{ ******** }
\message{       You'll have to answer a few questions in lowercase.}
\message{>  Do you want it in double-page (d), reduced (r)}
\message{or standard format (s) ? }\read-1 to\risposta
\message{>  Do you want it in USA A4 (u) or EUROPEAN A4 (e)}
\message{paper size ? }\read-1 to\srisposta
\message{>  Do you have AMS (math) fonts (y/n) ? }\read-1 to\arisposta
%
%
%
%
%
\ifx\risposta\standardrisposta
\message{>> This will come out UNREDUCED << }
\ingrandimento=1200 \dimnota=2 \unoduecol=1 \infralinea=16pt
\global\controllorisposta=1 \fi
\ifx\risposta\reducedrisposta
\ingrandimento=1095 \dimnota=1 \unoduecol=1 \infralinea=14pt
\message{>> This will come out REDUCED << }
\global\controllorisposta=1 \fi
\ifx\risposta\doublerisposta
\ingrandimento=1000 \dimnota=2 \unoduecol=2 \infralinea=12pt
\message{>> You must print this in LANDSCAPE orientation << }
\global\controllorisposta=1 \fi
\ifnum\controllorisposta=0
\message{>>> ERROR IN INPUT, I ASSUME STANDARD UNREDUCED FORMAT <<< }
\ingrandimento=1200 \dimnota=2 \unoduecol=1 \infralinea=16pt\fi
\def\magnificazione{\magnification=\ingrandimento}  \magnificazione
\ifx\risposta\doublerisposta
\ifx\srisposta\cartarisposta
\tothsize=25.5truecm \collhsize=12.0truecm \vsize=17.0truecm \else
\tothsize=9.4truein \collhsize=4.4truein \vsize=6.8truein \fi \else
\ifx\srisposta\cartarisposta
\tothsize=6.5truein \vsize=24truecm \else
\tothsize=6.5truein \vsize=8.9truein \fi
\collhsize=4.4truein \fi
\tolerance=10000
\leftskip=0pt \rightskip=0pt \sinnota=1 \parskip 0pt plus 2pt
%
%
%
%
\font\ninerm=cmr9  \font\eightrm=cmr8  \font\sixrm=cmr6
\font\ninei=cmmi9  \font\eighti=cmmi8  \font\sixi=cmmi6
\font\ninesy=cmsy9  \font\eightsy=cmsy8  \font\sixsy=cmsy6
\font\ninebf=cmbx9  \font\eightbf=cmbx8  \font\sixbf=cmbx6
\font\ninett=cmtt9  \font\eighttt=cmtt8 \font\nineit=cmti9
\font\eightit=cmti8 \font\ninesl=cmsl9  \font\eightsl=cmsl8
\skewchar\ninei='177 \skewchar\eighti='177 \skewchar\sixi='177
\skewchar\ninesy='60 \skewchar\eightsy='60 \skewchar\sixsy='60
\hyphenchar\ninett=-1 \hyphenchar\eighttt=-1 \hyphenchar\tentt=-1
%
\font\tencaps=cmcsc10              
%
\ifx\arisposta\amsrisposta
\font\mmmath=msym10                
\else \gdef\mmmath{\bf}  \fi
\ifnum\ingrandimento=1095

\font\capsone=cmcsc10 at 10.95pt 

\else

\font\capsone=cmcsc10 at 12pt 
\fi
\def\ttaarr{\bf}                
\def\ppaarr{\sl}                
\catcode`@=11
\newskip\ttglue
\gdef\tenpoint{\def\rm{\fam0\tenrm}
  \textfont0=\tenrm \scriptfont0=\sevenrm \scriptscriptfont0=\fiverm
  \textfont1=\teni \scriptfont1=\seveni \scriptscriptfont1=\fivei
  \textfont2=\tensy \scriptfont2=\sevensy \scriptscriptfont2=\fivesy
  \textfont3=\tenex \scriptfont3=\tenex \scriptscriptfont3=\tenex
  \textfont\itfam=\tenit \def\it{\fam\itfam\tenit}
  \textfont\slfam=\tensl \def\sl{\fam\slfam\tensl}
  \textfont\ttfam=\tentt \def\tt{\fam\ttfam\tentt}
  \textfont\bffam=\tenbf \scriptfont\bffam=\sevenbf
  \scriptscriptfont\bffam=\fivebf \def\bf{\fam\bffam\tenbf}
  \tt \ttglue=.5em plus.25em minus.15em
  \normalbaselineskip=12pt
  \setbox\strutbox=\hbox{\vrule height8.5pt depth3.5pt width0pt}
  \let\sc=\eightrm \let\big=\tenbig \normalbaselines\rm}
\gdef\ninepoint{\def\rm{\fam0\ninerm}
  \textfont0=\ninerm \scriptfont0=\sixrm \scriptscriptfont0=\fiverm
  \textfont1=\ninei \scriptfont1=\sixi \scriptscriptfont1=\fivei
  \textfont2=\ninesy \scriptfont2=\sixsy \scriptscriptfont2=\fivesy
  \textfont3=\tenex \scriptfont3=\tenex \scriptscriptfont3=\tenex
  \textfont\itfam=\nineit \def\it{\fam\itfam\nineit}
  \textfont\slfam=\ninesl \def\sl{\fam\slfam\ninesl}
  \textfont\ttfam=\ninett \def\tt{\fam\ttfam\ninett}
  \textfont\bffam=\ninebf \scriptfont\bffam=\sixbf
  \scriptscriptfont\bffam=\fivebf \def\bf{\fam\bffam\ninebf}
  \tt \ttglue=.5em plus.25em minus.15em
  \normalbaselineskip=11pt
  \setbox\strutbox=\hbox{\vrule height8pt depth3pt width0pt}
  \let\sc=\sevenrm \let\big=\ninebig \normalbaselines\rm}
\gdef\eightpoint{\def\rm{\fam0\eightrm}
  \textfont0=\eightrm \scriptfont0=\sixrm \scriptscriptfont0=\fiverm
  \textfont1=\eighti \scriptfont1=\sixi \scriptscriptfont1=\fivei
  \textfont2=\eightsy \scriptfont2=\sixsy \scriptscriptfont2=\fivesy
  \textfont3=\tenex \scriptfont3=\tenex \scriptscriptfont3=\tenex
  \textfont\itfam=\eightit \def\it{\fam\itfam\eightit}
  \textfont\slfam=\eightsl \def\sl{\fam\slfam\eightsl}
  \textfont\ttfam=\eighttt \def\tt{\fam\ttfam\eighttt}
  \textfont\bffam=\eightbf \scriptfont\bffam=\sixbf
  \scriptscriptfont\bffam=\fivebf \def\bf{\fam\bffam\eightbf}
  \tt \ttglue=.5em plus.25em minus.15em
  \normalbaselineskip=9pt
  \setbox\strutbox=\hbox{\vrule height7pt depth2pt width0pt}
  \let\sc=\sixrm \let\big=\eightbig \normalbaselines\rm}
\gdef\tenbig#1{{\hbox{$\left#1\vbox to8.5pt{}\right.\n@space$}}}
\gdef\ninebig#1{{\hbox{$\textfont0=\tenrm\textfont2=\tensy
   \left#1\vbox to7.25pt{}\right.\n@space$}}}
\gdef\eightbig#1{{\hbox{$\textfont0=\ninerm\textfont2=\ninesy
   \left#1\vbox to6.5pt{}\right.\n@space$}}}
%
\newbox\fotlinebb \newbox\hedlinebb \newbox\leftcolumn
\gdef\makeheadline{\vbox to 0pt{\vskip-22.5pt
     \fullline{\vbox to8.5pt{}\the\headline}\vss}\nointerlineskip}
\gdef\makehedlinebb{\vbox to 0pt{\vskip-22.5pt
     \fullline{\vbox to8.5pt{}\copy\hedlinebb\hfil
     \line{\hfill\the\headline\hfill}}\vss}
     \nointerlineskip}
\gdef\makefootline{\baselineskip=24pt \fullline{\the\footline}}
\gdef\makefotlinebb{\baselineskip=24pt
    \fullline{\copy\fotlinebb\hfil\line{\hfill\the\footline\hfill}}}
\gdef\doubleformat{\shipout\vbox{\makehedlinebb
     \fullline{\box\leftcolumn\hfil\columnbox}\makefotlinebb}
     \advancepageno}

\gdef\columnbox{\leftline{\pagebody}}
\gdef\line#1{\hbox to\hsize{\hskip\leftskip#1\hskip\rightskip}}
\gdef\fullline#1{\hbox to\fullhsize{\hskip\leftskip{#1}%
\hskip\rightskip}}
\gdef\footnote#1{\let\@sf=\empty
         \ifhmode\edef\#sf{\spacefactor=\the\spacefactor}\/\fi
         #1\@sf\vfootnote{#1}}
\gdef\vfootnote#1{\insert\footins\bgroup
         \ifnum\dimnota=1  \eightpoint\fi
         \ifnum\dimnota=2  \ninepoint\fi
         \ifnum\dimnota=0  \tenpoint\fi
         \interlinepenalty=\interfootnotelinepenalty
         \splittopskip=\ht\strutbox
         \splitmaxdepth=\dp\strutbox \floatingpenalty=20000
         \leftskip=\oldssposta \rightskip=\olddsposta
         \spaceskip=0pt \xspaceskip=0pt
         \ifnum\sinnota=0   \textindent{#1}\fi
         \ifnum\sinnota=1   \item{#1}\fi
         \footstrut\futurelet\next\fo@t}
\gdef\fo@t{\ifcat\bgroup\noexpand\next \let\next\f@@t
             \else\let\next\f@t\fi \next}
\gdef\f@@t{\bgroup\aftergroup\@foot\let\next}
\gdef\f@t#1{#1\@foot}
\gdef\@foot{\strut\egroup}
\gdef\footstrut{\vbox to\splittopskip{}}
\skip\footins=\bigskipamount
\count\footins=1000  \dimen\footins=8in
\catcode`@=12
\tenpoint  \baselineskip=\infralinea
\newskip\olddsposta \newskip\oldssposta
\global\oldssposta=\leftskip \global\olddsposta=\rightskip

\gdef\yespagenumbers{\footline={\hss\tenrm\folio\hss}}
\gdef\ciao{\par\vfill\supereject
      \ifnum\unoduecol=2 \if R\lrcol \null\vfill\eject \fi\fi \end}

\ifnum\unoduecol=1 \hsize=\tothsize   \fullhsize=\tothsize \fi
\ifnum\unoduecol=2 \hsize=\collhsize  \fullhsize=\tothsize \fi
\global\let\lrcol=L
\ifnum\unoduecol=1 \output{\plainoutput}\fi
\ifnum\unoduecol=2 \output{\if L\lrcol
       \global\setbox\leftcolumn=\columnbox
       \global\setbox\fotlinebb=\line{\hfill\the\footline\hfill}
       \global\setbox\hedlinebb=\line{\hfill\the\headline\hfill}
       \advancepageno
      \global\let\lrcol=R \else \doubleformat \global\let\lrcol=L \fi
       \ifnum\outputpenalty>-20000 \else\dosupereject\fi}\fi
\def\ifdoublepage{\ifnum\unoduecol=2 }
\def\filldots{\leaders\hbox to 1em{\hss.\hss}\hfill}
\def\inquadrb#1 {\vbox {\hrule  \hbox{\vrule \vbox {\vskip .2cm
    \hbox {\ #1\ } \vskip .2cm } \vrule  }  \hrule} }

\def\newline{\hfil\break}
\def\jump{\vskip\baselineskip} \newskip\iinnffrr
\def\sjump{\iinnffrr=\baselineskip
          \divide\iinnffrr by 2 \vskip\iinnffrr}
\def\bjump{\vskip\baselineskip \vskip\baselineskip}
\newcount\nmbnota  \def\clearnmbnota{\global\nmbnota=0}
\def\note#1{\global\advance\nmbnota by 1
    \footnote{$^{\the\nmbnota}$}{#1}}  \clearnmbnota
\def\setnote#1{\expandafter\xdef\csname#1\endcsname{\the\nmbnota}}
\newcount\nbmfig  \def\clearnbmfig{\global\nbmfig=0}
\gdef\figure{\global\advance\nbmfig by 1
      {\rm fig. \the\nbmfig}}   \clearnbmfig
\def\setfig#1{\expandafter\xdef\csname#1\endcsname{fig. \the\nbmfig}}
 \def\endformula{\eqno\numero $$}
 \def\efr{\endformula}
\newcount\frmcount \def\clearfrmcount{\global\frmcount=0}
\def\numero{\global\advance\frmcount by 1   \ifnum\indappcount=0
  {\ifnum\cpcount <1 {\hbox{\rm (\the\frmcount )}}  \else
  {\hbox{\rm (\the\cpcount .\the\frmcount )}} \fi}  \else
  {\hbox{\rm (\applett .\the\frmcount )}} \fi}
\def\nameformula#1{\global\advance\frmcount by 1%
\ifnum\draftnum=0  {\ifnum\indappcount=0%
{\ifnum\cpcount<1\xdef\spzzttrra{(\the\frmcount )}%
\else\xdef\spzzttrra{(\the\cpcount .\the\frmcount )}\fi}%
\else\xdef\spzzttrra{(\applett .\the\frmcount )}\fi}%
\else\xdef\spzzttrra{(#1)}\fi%
\expandafter\xdef\csname#1\endcsname{\spzzttrra}
\eqno \ifnum\draftnum=0 {\ifnum\indappcount=0
  {\ifnum\cpcount <1 {\hbox{\rm (\the\frmcount )}}  \else
  {\hbox{\rm (\the\cpcount .\the\frmcount )}}\fi}   \else
  {\hbox{\rm (\applett .\the\frmcount )}} \fi} \else (#1) \fi $$}
\def\nfr{\nameformula}    
\def\nameali#1{\global\advance\frmcount by 1%
\ifnum\draftnum=0  {\ifnum\indappcount=0%
{\ifnum\cpcount<1\xdef\spzzttrra{(\the\frmcount )}%
\else\xdef\spzzttrra{(\the\cpcount .\the\frmcount )}\fi}%
\else\xdef\spzzttrra{(\applett .\the\frmcount )}\fi}%
\else\xdef\spzzttrra{(#1)}\fi%
\expandafter\xdef\csname#1\endcsname{\spzzttrra}
\eqno \ifnum\draftnum=0  {\ifnum\indappcount=0
  {\ifnum\cpcount <1 {\hbox{\rm (\the\frmcount )}}  \else
  {\hbox{\rm (\the\cpcount .\the\frmcount )}}\fi}   \else
  {\hbox{\rm (\applett .\the\frmcount )}} \fi} \else (#1) \fi}
\clearfrmcount
\newcount\cpcount 
\newcount\subcpcount \def\clearsubcpcount{\global\subcpcount=0}
\newcount\appcount \def\clearappcount{\global\appcount=0}
\newcount\indappcount \def\clearindappcount{\indappcount=0}
\newcount\sottoparcount 

\def\applett{\ifcase\appcount  \or {A}\or {B}\or {C}\or
{D}\or {E}\or {F}\or {G}\or {H}\or {I}\or {J}\or {K}\or {L}\or
{M}\or {N}\or {O}\or {P}\or {Q}\or {R}\or {S}\or {T}\or {U}\or
{V}\or {W}\or {X}\or {Y}\or {Z}\fi
             \ifnum\appcount<0
    \message{>>  ERROR: counter \appcount out of range <<}\fi
             \ifnum\appcount>26
   \message{>>  ERROR: counter \appcount out of range <<}\fi}
\clearappcount  \clearindappcount
\newcount\connttrre  \def\clearconnttrre{\global\connttrre=0}
\newcount\countref  \def\clearcountref{\global\countref=0}
\clearcountref
\def\chapter#1{\global\advance\cpcount by 1 \clearfrmcount
                 \goodbreak\null\jump\nobreak
                 \clearsubcpcount\clearindappcount
                 \itemitem{\ttaarr\the\cpcount .\qquad}{\ttaarr #1}
                 \par\nobreak\jump\sjump\nobreak}
\def\section#1{\global\advance\subcpcount by 1 \goodbreak\null
                  \sjump\nobreak\ifnum\indappcount=0
                 {\ifnum\cpcount=0 {\itemitem{\ppaarr
               .\the\subcpcount\quad\enskip\ }{\ppaarr #1}\par} \else
                 {\itemitem{\ppaarr\the\cpcount .\the\subcpcount\quad
                  \enskip\ }{\ppaarr #1} \par}  \fi}
                \else{\itemitem{\ppaarr\applett .\the\subcpcount\quad
                 \enskip\ }{\ppaarr #1}\par}\fi\nobreak\jump\nobreak}
\clearsubcpcount
\def\appendix#1{\global\advance\appcount by 1 \clearfrmcount
                  \goodbreak\null\jump\nobreak
                  \global\advance\indappcount by 1 \clearsubcpcount
                  \itemitem{\ttaarr App.\applett\ }{\ttaarr #1}
                  \nobreak\jump\sjump\nobreak}
\clearappcount \clearindappcount
\def\references{\goodbreak\null\jump\nobreak
   \itemitem{}{\ttaarr References} \nobreak\jump\sjump\nobreak}
\clearcountref

\def\acknowledgements{\goodbreak\null\jump\nobreak
\itemitem{ }{\ttaarr Acknowledgements} \nobreak\jump\sjump\nobreak}
\def\setchap#1{\ifnum\indappcount=0{\ifnum\subcpcount=0%
\xdef\spzzttrra{\the\cpcount}%
\else\xdef\spzzttrra{\the\cpcount .\the\subcpcount}\fi}
\else{\ifnum\subcpcount=0 \xdef\spzzttrra{\applett}%
\else\xdef\spzzttrra{\applett .\the\subcpcount}\fi}\fi
\expandafter\xdef\csname#1\endcsname{\spzzttrra}}
    \newcount\draftnum
\newcount\ppora   \newcount\ppminuti
\global\ppora=\time   \global\ppminuti=\time
\global\divide\ppora by 60  \draftnum=\ppora
\multiply\draftnum by 60    \global\advance\ppminuti by -\draftnum
\global\draftnum=0
\def\droggi{\number\day /\number\month /\number\year\ \the\ppora
:\the\ppminuti}

\global\draftnum=0
%
%
\catcode`@=11
\gdef\Ref#1{\expandafter\ifx\csname @rrxx@#1\endcsname\relax%
{\global\advance\countref by 1%
\ifnum\countref>200%
\message{>>> ERROR: maximum number of references exceeded <<<}%
\expandafter\xdef\csname @rrxx@#1\endcsname{0}\else%
\expandafter\xdef\csname @rrxx@#1\endcsname{\the\countref}\fi}\fi%
\ifnum\draftnum=0 \csname @rrxx@#1\endcsname \else#1\fi}
\gdef\beginref{\ifnum\draftnum=0  \gdef\Rref{\fairef}
\gdef\endref{\scriviref} \else\relax\fi}
\gdef\Rref#1#2{\item{[#1]}{#2}}  \gdef\endref{\relax}
\newcount\conttemp
\gdef\fairef#1#2{\expandafter\ifx\csname @rrxx@#1\endcsname\relax
{\global\conttemp=0
\message{>>> ERROR: reference [#1] not defined <<<} } \else
{\global\conttemp=\csname @rrxx@#1\endcsname } \fi
\global\advance\conttemp by 50
\global\setbox\conttemp=\hbox{#2} }
\gdef\scriviref{\clearconnttrre\conttemp=50
\loop\ifnum\connttrre<\countref \advance\conttemp by 1
\advance\connttrre by 1
\item{[\the\connttrre]}{\unhcopy\conttemp} \repeat}
\clearcountref \clearconnttrre
\catcode`@=12
\def\slashchar#1{\setbox0=\hbox{$#1$} \dimen0=\wd0
     \setbox1=\hbox{/} \dimen1=\wd1 \ifdim\dimen0>\dimen1
      \rlap{\hbox to \dimen0{\hfil/\hfil}} #1 \else
      \rlap{\hbox to \dimen1{\hfil$#1$\hfil}} / \fi}
\ifx\oldchi\undefined \let\oldchi=\chi
  \def\cchi{{\raise 1pt\hbox{$\oldchi$}}} \let\chi=\cchi \fi

\def\frac#1#2{{\textstyle{#1 \over #2}}}

\def\half{\ifinner {\scriptstyle {1 \over 2}}\else {1 \over 2} \fi}

\def\simge{\rlap{\raise 2pt \hbox{$>$}}{\lower 2pt \hbox{$\sim$}}}
\def\simle{\rlap{\raise 2pt \hbox{$<$}}{\lower 2pt \hbox{$\sim$}}}

\def\vbig#1#2{{\vbigd@men=#2\divide\vbigd@men by 2%
\hbox{$\left#1\vbox to \vbigd@men{}\right.\n@space$}}}

\null
%
%
%
%

\nopagenumbers{\baselineskip=12pt
\line{\hfill IASSNS-HEP-91/61}
\line{\hfill PUPT-1285}
\line{\hfill September, 1991}
\ifdoublepage \bjump\bjump\else\vfill\fi
\centerline{\capsone GENERALIZED W-ALGEBRAS AND INTEGRABLE}
\sjump
\centerline{\capsone HIERARCHIES}
\bjump\bjump
\centerline{\tencaps Nigel Burroughs \& Mark de Groot}
\sjump
\centerline{\sl Institute for Advanced Study,}
\centerline{\sl Olden Lane, Princeton, N.J. 08540.}
\bjump
\centerline{\tencaps Timothy Hollowood}
\sjump
\centerline{Dept. of Theoretical Physics,}
\centerline{Oxford, U.K. OX1 3NP}
\bjump
\centerline{\tencaps Luis Miramontes}
\sjump
\centerline{\sl Theory Division, CERN, 1211-Geneva 23,}
\centerline{\sl Switzerland.}
\bjump
\vfill
\ifnum\unoduecol=2 \eject\null\vfill\fi
\centerline{\capsone ABSTRACT}
\sjump
\noindent
We report on generalizations of the KdV-type integrable hierarchies of
Drinfel'd
and Sokolov. These  hierarchies lead to the existence of new
classical $W$-algebras, which  arise as the second Hamiltonian
structure of the hierarchies. In particular,
we present a construction of the $W_n^{(l)}$ algebras.
\sjump
\ifnum\unoduecol=2 \vfill\fi
\eject}
\yespagenumbers\pageno=1

%
Several years ago, Drinfel'd and Sokolov obtained a very significant
generalization of the usual Korteweg-de Vries (KdV)
hierarchy [\Ref{Sc}]. This hierarchy, the first non-trivial flow of which is
$$
{\partial u\over\partial {t_1}}=-{1\over 4}u^{\prime\prime\prime}+
{3\over 2}uu^\prime\
 {\rm where} \ u^{\prime}\equiv{\partial u\over\partial x},
\efr
can be
straightforwardly obtained in terms of pseudo-differential operators
[\Ref{Ri}]. To this end one considers the scalar Lax equation
$$
{dL\over dt}=[A,L],\ {\rm where}\
 L=D^k+\sum_{i=1}^{k}u_iD^{i-1},\ A=\sum_{i=1}^{m+1}v_iD^{i-1}.
\nfr{slax}
It can be shown that
$A$ is determined by $L$, up to $m$ arbitrary constants and and one
arbitrary function;
also we note that since the right-hand side of the equation can be shown to
have
order less than
 or equal to $k-2$ we can consistently set $u_{k-1}=0$.
The case $\ k=2 \ $ then corresponds to the usual KdV equation.

Now, these scalar Lax equations above can be given a matrix
representation without  difficulty [\Ref{Sc}]. Choosing
$$
\tilde L=\partial_x+q+\Lambda=
\partial_x+\pmatrix{0&\cdots&0&0\cr  \vdots&&\vdots&\vdots\cr
0&\cdots&0&0\cr u_1&\cdots&u_n&0\cr}+\pmatrix{&1&&\cr &&\ddots&\cr
&&&1\cr z&&&\cr}
\nfr{mlax}
with $z$ a spectral parameter,  the scalar Lax equation
 becomes the matrix
Lax equation
$$
{d\tilde L\over dt}=\left[\tilde A,\tilde L\right],
\efr
where the differential operator $A$, of order $m$, becomes a matrix
$\tilde A$ of the form $\tilde A=\sum_{i=0}^mz^i\tilde v_i$. (This corresponds
to  $\ k=n+1\ $ in \slax, where we have also set $u_{n+1}=0$ as allowed.)

A natural question is whether the above construction can be generalized. One
could  consider a more general $q$, for example allow it to be
a lower triangular
matrix. In this case, though, one discovers  that  the matrix $\tilde A$ whose
commutator  with $\tilde L$ would give the evolution equation of the
hierarchy at a certain order, is not uniquely defined.
However, introducing
the  notion
of gauge-equivalence
$\tilde L\rightarrow\tilde L'=N\tilde L N^{-1}$
with $N$  a
lower-triangular matrix with 1's on the diagonal, allows one to
circumvent this indeterminacy and by choosing an appropriate gauge-slice one
can actually
re-obtain the scalar Lax equation. The advantage of  this somewhat
complicated rewriting of the scalar Lax equation is that whereas the $q$ in
equation \mlax\ has
no group-theoretic interpretation, the lower-triangular $q$ can be considered
an element of the Borel subalgebra, while the  $N$ can be considered an
exponentiated  subalgebra, of $\ sl(n+1)\ $. In fact our
earlier KdV equations are now seen to
correspond to  $\ sl(n+1,{\mmmath C} [z,z^{-1}])\ $ Kac-Moody algebras;
this interpretation allows a natural
generalization, and so  KdV type equations can be defined for arbitrary
Kac-Moody algebras [\Ref{Sc}].

Similar results were obtained
in [\Ref{Si}], though the association  here was made between {\it  modified\/}
KdV equations and untwisted
affine Lie algebras. In this same work a suggestion was made concerning the
proper setting in which these developments  were to be viewed
and it was this suggestion that provided an impetus for our work.
Apart from their
being examples of new integrable systems, the new hierarchies are also
of interest because of they lead to generalized $W$-algebras.
Essentially, systems of the KdV-type
admit two distinct, but {\it coordinated\/}, Hamiltonian structures. The
`second' Poisson bracket algebra is then the $W$-algebra associated to
the hierarchy: for example, the $sl(n+1)$ hierarchies defined above,
via \mlax, lead to the $W_{n+1}$-algebras, where the $u_i$'s are the
generators. Our construction includes the $W_3^{(2)}$-algebra [\Ref{Sg}],
which is related  to an example of a `fractional' KdV hierarchy [\Ref{Se}].
The generalized hierarchies and $W$-algebras that we obtain appear to be
related to other approaches [\Ref{genws}].

We now explain the general construction, a more complete treatment
appearing in refs. [\Ref{Rg},\Ref{Bd}].
The central object in the construction will be the loop algebra
associated to a finite Lie algebra, $\hat g=g\otimes{\mmmath
C}[z,z^{-1}]$, equipped with a
derivation $d=z(d/dz)$, which induces the {\it homogeneous
gradation\/}. Other gradations may be introduced by choosing
${\rm rank}(g)+1$ non-negative integers ${\bf s}=(s_0,\ldots,s_{{\rm
rank}(g)})$, and defining the derivation [\Ref{Rb}]
$$
d_{\bf s}=N_{\bf s}d+\sum_{k=1}^{{\rm
rank}(g)}\left({2\over\alpha_k^2}\right)s_k\omega_k\cdot H,
\efr
where $N_{\bf s}=\sum_{i=0}^{{\rm rank}(g)}k_is_i$ (the $k_i$ being the
Kac-labels of $g$), the $\omega_i$ are the fundamental weights of $g$,
and $H$ is the Cartan subalgebra of $g$. The derivation $d_{\bf s}$
then defines the ${\bf s}$-gradation: $\hat g=\bigoplus_{i\in{\bf
Z}}\hat g_i({\bf s})$.

For later use define
$ \hat g_{>i}({\bf s})=\bigoplus_{k>i}\hat g_k({\bf s}) $
and similarly for $\ \hat g_{<i}({\bf s})$, $\hat g_{\leq i}({\bf s})$
and $\hat g_{\geq i}({\bf s})$; also we say that ${\bf s}\succeq{\bf
s}^\prime$ provided that $s_i\neq0$ whenever $s_i^\prime\neq0$, and we
thereby obtain a partial ordering on the set of gradations $\{{\bf s}\}$
of $\hat g$.
For convenience we shall  use superscripts to denote ${\bf
s}[w]$-grades and subscripts to denote ${\bf s}$ grades.
To simplify notation we assume that we
are working in some faithful matrix representation of $g$.

There is a class of subalgebras of $\hat g$ which will
play an important r\^ole in what follows, these being the {\it
Heisenberg subalgebras\/}. They are special types of maximally commuting
subalgebras whose precise definition may be found in [\Ref{Sm}], in which it is
further shown
that the inequivalent Heisenberg
subalgebras of $\hat g$ are
classified by the conjugacy classes of the Weyl group of $g$. We
denote the Heisenberg subalgebra corresponding to the conjugacy class
containing the Weyl reflection $w$, as ${\cal H}[w]$. An
important property of these algebras
is that they are associated to a particular gradation of
$\hat g$, which we denote ${\bf s}[w]$, such that ${\cal H}[w]$ is an
invariant subspace of ad$\,d_{{\bf s}[w]}$.
 At this point we make the
parenthetic remark, that for the following construction  we need
 a particular gradation, and we then need  to be able to choose an
element which has a well-defined grade and is semi-simple,
i.e. one such that the Kac-Moody algebra  $\hat g$ equals the direct
sum of the kernel plus the image of the action of this element on $\hat g$.
The procedure
outlined above, where we lift conjugacy classes of the Weyl group, and look
at the associated gradations, is a means of accomplishing this in a systematic
fashion.

We begin by defining a differential operator
$\ L\equiv\partial_x+q+\Lambda\ $,
associated to the data $(\Lambda,w,{\bf s})$ , in the space
$C^\infty({\mmmath R},\hat g)$ , where $\hat g$ is the loop algebra of a
finite Lie algebra $g$. The element $\Lambda$ is a  regular element of ${\cal
H}[w]$ which is constant
($\partial_x\Lambda=0$) with positive ${\bf s}[w]$-grade $i$, where $w$ is some
   element
of the Weyl group of $g$ --- conjugate elements
of the Weyl group leading  to isomorphic hierarchies.
Here by regular we mean that Ker$({\rm ad}\,\Lambda)={\cal H}[w]$.
(If this latter condition is not satisfied, one can still define
a hierarchy but a bit more care must be taken [\Ref{Si},\Ref{Rg}].)
Lastly the {\it potential\/}
$q\in C^\infty({\mmmath R},Q)$, where $Q$ is defined
to be the following subspace of $\hat g$,
$\ Q=\hat g_{\geq0}({\bf s})\cap\hat g^{<i}({\bf s}[w])\ $,
${\bf s}$  some other
gradation of $\hat g$ such that ${\bf s}\preceq{\bf s}[w]$.

Having imposed the requisite conditions on the terms in $L$ we now
assert that there exist an infinite number of commuting
flows of the form
$$
{\partial L\over\partial t_b}=[A(b),L].
\nfr{BAA}
Here $b$ is a constant element of the subspace of
${\cal H}[w]$ with positive ${\bf s}[w]$-grade, and
$A(b)\in C^\infty({\mmmath R},\hat g)$.

To prove this,  define the following exponentiated action
of the loop algebra on $L$, ${\cal L}=\Phi L\Phi^{-1}$.
It can be shown that  there exists a (non-unique) transformation of
this type, for which ${\cal L}$ has the form
$$
{\cal L}=\partial_x+\Lambda+\sum_{j<i}h^j,
\nfr{BC}
where $h^j\in C^\infty({\mmmath R},{\cal H}^j[w])$.
If we then have an $M$ such that $[M,L]=0$, setting
 $A=M_{\geq0}$ or $A=M^{\geq0}$, we find that that
 $[A,L]\in C^{\infty}({\mmmath R},Q)$, see [\Ref{Rg}].
Since a constant element of ${\cal H}[w]$, say $b$, commutes with ${\cal
L}$, the quantity $\Phi^{-1}b\Phi$ commutes with $L$. Using the above
result we deduce that two sets of flows of the form \BAA\ exist with
$A(b)=(\Phi^{-1}b\Phi)_{\geq0}$ or $A(b)=(\Phi^{-1}b\Phi)^{\geq0}$,
respectively.

We now introduce the key concept of a gauge symmetry.
The form of $L$ is preserved under  the transformation
 $L^\prime=SLS^{-1}$, $S$ being generated by the subalgebra of
functions on
$P=\hat g_0\cap\hat g^{<0}$. Under this action:
$$
q\longmapsto S(\partial_x+q+\Lambda)S^{-1}-\partial_x-\Lambda.
\efr
The phase space of the hierarchy is actually equal to
$C^\infty({\mmmath R},Q)$ modulo the action of these gauge transformations.

We can  write consistent flows for a gauge fixed operator $\tilde
L$, by choosing $\tilde q\in C^\infty({\bf R},\tilde Q)$, where
$\tilde Q$ is some consistent gauge slice.
However, if we do this then there is
no guarantee that the flows \BAA\  will preserve the gauge slice chosen. To
compensate for this it will be necessary to modify the flows by
 gauge transformations:
$$
{\partial\tilde L\over\partial t}=\left[\left(\Phi^{-1}b\Phi
\right)_{\geq0}+\theta,\tilde L\right]\ {\rm and }\ \
{\partial\tilde L\over\partial t^\prime}=\left[\left(\Phi^{-1}b
\Phi\right)^{\geq0}+\theta^\prime,\tilde L\right],
\nfr{BL}
for some $\theta,\theta^\prime\in C^\infty({\mmmath R},P)$.
It can be demonstrated that the equations in \BL\ lead to  the same
evolution equation for $\tilde q$, and also that
the quantities $h^j$ of \BC\ are the conserved densities for the hierarchy.
In addition, the finite set of $h^j$'s with $i>j\geq0$ are
constant along all the flows (recall that $i$ is defined by the fact that
$\Lambda\in{\cal H}^i[w]$), {\it i.e.\/} for any flow
$$
\partial_th^j=0\ \ \ i>j\geq0.
\efr
This result means that some components of $q$ can be consistently
set to zero, since they are constant under the flows.

We can consider the spectrum of hierarchies for fixed $w$ and
$\Lambda$ but allowing the gradation $\bf s$ to vary ---  it has only to
satisfy
${\bf s}\preceq{\bf s}[w]$. For ${\bf s}={\bf s}[w]$ it is apparent
from the definition of $P$ that there is
no gauge invariance and so we have a generalization of the
Drinfel'd-Sokolov {\it modified\/}-KdV hierarchies. On the contrary,
when ${\bf s}$ is as `small' as it can be, for instance
the homogeneous gradation,
the gauge invariance is maximal, and then we have a generalization
of the Drinfel'd-Sokolov KdV-type hierarchies. Choosing a gradation ${\bf
s}$ between these two extremes leads to what we call a {\it
partially modified\/} KdV hierarchy (pmKdV). It turns out that all
these hierarchies are related by a series of Miura maps [\Ref{Bd}].

The hierarchies constructed by Drinfel'd and Sokolov are recovered
from our formalism by choosing $[w]$ to be the Coxeter class $[w_c]$, and
taking $\Lambda$ to have ${\bf s}[w]$-grade 1. There
are two ways to generalize these known cases. Firstly,
we can
take $\Lambda$ to be a different element of the Heisenberg
subalgebra corresponding to the Coxeter class, and secondly
 we can choose
$\Lambda$ to be an element of a
new conjugacy class of the Weyl group.
The former generalization leads to the fractional KdV
hierarchies [\Ref{Se}].

As an example of the above analysis,
we consider the fractional KdV hierarchies for the
$sl(n+1)$-algebras. For
$g=sl(n+1)$, the Heisenberg subalgebra ${\cal H}[w_c]$ is spanned
by the elements
$$
\Lambda_{j,m}=z^m\pmatrix{{\bf0}&{\bf I}_{n+1-j}\cr z{\bf
I}_j&{\bf0}\cr}\ \ \ j=1,2,\ldots,n,\ \
m\in{\bf Z},
\nfr{CB}
${\bf I}_i$ being  the $i\times i$ unit matrix.
So for a given $n$ we have a whole tower of hierarchies, for which
$\Lambda=\Lambda_{j,m}$ $m\geq0$; the
original hierarchy of Drinfel'd and Sokolov being at the bottom.
If we consider  $sl(2)$ and
$i=3$, we find that the gauge-fixed form
differs from that of the usual KdV form, explicitly
$$
\tilde q=\pmatrix{\tilde a&-\tilde c\cr \tilde b+\tilde cz&-\tilde a}\
 {\rm as\ compared\ to\ } \pmatrix{0&0\cr a&0\cr}.
\efr
Proceeding as outlined above,
we find that the equations of motion for the first and second
flow of the fractional  hierarchy are
$$
\partial_{\tilde t_1}\pmatrix{\tilde a&-\tilde c\cr\tilde b+\tilde c z&-\tilde
    a\cr}
=\pmatrix{\tilde b+2\tilde c^{2}&-2\tilde a\cr2\tilde az-2\partial_{\tilde x}
\tilde c+4\tilde c
\tilde a&-\tilde b-2\tilde c^2\cr}
\efr
for the first,
while the second is trivial ( it  identifies
$t_3$ with $-x$) .

We now turn to a discussion of the canonical formalism for the
hierarchies.
A feature often encountered in the Hamiltonian analysis of integrable
hierarchies, is the presence
of two {\it coordinated\/} Poisson structures which we designate
$\{\phi,\psi\}_1$
and $\{\phi,\psi\}_2$. The property of {\it coordination\/} implies
that the one-parameter family of brackets
$$
\{\phi,\psi\}=\{\phi,\psi\}_1+\mu\{\phi,\psi\}_2,
\efr
$\mu$ arbitrary, is also a Poisson structure, which is a non-trivial statement
as regards the Jacobi
identity. We say a system has a {\it bi-Hamiltonian
structure\/} if the brackets are coordinated and if the Hamiltonian flow can be
written in
two equivalent ways
$$
{\dot\phi}=\{H_2,\phi\}_1=\{H_1,\phi\}_2.
\efr
Under various general assumptions, the existence of a bi-Hamiltonian
structure implies the existence of an infinite hierarchy of flows,
that is, an infinite set of Hamiltonians $\{H_i\}$, such that
$$\partial_{t_i}\phi=\{H_{i+1},\phi\}_1=\{H_i,\phi\}_2,$$
where the Hamiltonians are in involution with respect to both
Poisson brackets, whence  the flows $\partial_{t_i}$ commute.

An example of these features is provided by the original KdV
hierarchy which
has the two Poisson structures
$$
\eqalign{\{u(x),u(y)\}_1&=2\delta^\prime(x-y)\cr
\{u(x),u(y)\}_2&={1\over2}\delta^{\prime\prime\prime
}(x-y)-2u(x)\delta^\prime(x-y)-u^\prime(x)\delta(x-y).\cr}
\nfr{kdv}
We can similarly obtain Poisson brackets for the hierarchies
introduced above and shall  find amongst other things that, in the same
way as  the second structure of the KdV
equation is a Virasoro algebra, so the second structures of the new
hierarchies constitute generalized W-algebras (for details see [\Ref{Bd}]).
There are  hierarchies that do not
admit a bi-Hamiltonian structure, for example the {\it modified\/} KdV
hierarchy (mKdV) and its generalizations.
However, it is known that the single Hamiltonian structure is the
pull-back of the second Hamiltonian structure of the associated KdV
hierarchy by the Miura map connecting the hierarchies. This property
extends to the generalized hierarchies; each KdV-type hierarchy admits
a bi-Hamiltonian structure, whilst the associated pmKdV and mKdV
hierarchies admit only a single Hamiltonian structure [\Ref{Bd}].

The function $q(x)$ plays the r\^ole of the phase space coordinate in
this system. However, as we saw earlier, there exist symmetries
in the system corresponding to the gauge transformations
$L\rightarrow SLS^{-1}.$
The {\it phase
space\/} of the system ${\cal M}$ is the set of gauge equivalence
classes of operators of the form $L=\partial_x+q+\Lambda$.
The space of functions ${\cal F}$ on ${\cal M}$ is the
set of gauge invariant
functionals of $q$ of the form
$$
\varphi[q]=\int_{{\mmmath S}^1}
dx\,f\left(x,q(x),q^\prime(x),\ldots,q^{(n)}(x),\ldots\right),
\efr
where we now take periodic functions for convenience (following
[\Ref{Sc}]).
It is straightforward to find a basis for ${\cal F}$, the gauge
invariant functionals. One simply performs a non-singular
gauge transformation to take $q$ to some canonical form $\tilde q$.
The coordinates $\tilde q$ and their derivatives then provide the desired
basis. For
instance, for the generalized $sl(n+1)$-KdV hierarchies of Drinfel'd and
Sokolov, $q$ consists of
lower triangular $n+1$ by $n+1$ dimensional matrices, while the gauge
group is generated by strictly lower triangular matrices. A good gauge slice
consists of matrices of the form of $q$ in \mlax,
the $u_i$'s and their
derivatives providing a basis for $\cal F$.

There is a
natural inner product on the functions $C^\infty({\mmmath S}^1,\hat g)$,
defined as follows
$$
(A,B)=\int_{{\mmmath S}^1}\ dx\,\langle A(x),B(x)
\rangle_{\hat g},
\efr
where $\langle\ ,\ \rangle_{\hat g}$ is the Killing form of $\hat g$.
Explicitly
$\langle a\otimes z^n,b\otimes z^m\rangle_{\hat g}=\langle
a,b\rangle_g\delta_{n+m,0},$
where $\langle\ ,\ \rangle_g$ is the Killing form of $g$.

We now proceed to define the two Poisson structures for the KdV
hierarchies (for which ${\bf s}$ is the homogeneous gradation).
For a constant element $b \in {\cal H}[w]^{>0}$, define
the following functional of $q$,
$H_b[q]=(b,h(q)),$
where $h(q)=\sum_{j<i}h^j$ is the sum of the conserved densities of
\BC.   For a functional $\varphi$ of $q$ define its
functional derivative $d_q\varphi\in C^\infty({\mmmath S}^1,\hat g_{\leq0})$
via
$$
\left.{d\over d\varepsilon}\varphi[q+\varepsilon r]\right\vert_{\varepsilon=0}
\equiv\left(d_q\varphi,r\right),\qquad\quad\forall r\in
C^\infty({\mmmath S}^1,Q).
\efr
Notice that there is an ambiguity
in the definition of the functional derivative $d_q\varphi$
corresponding to the fact that terms in the annihilator of $Q$ are
not determined; despite this our subsequent definitions
are well-defined [\Ref{Bd}].
The two coordinated Hamiltonian structures on ${\cal F}$ are
$$\eqalign{
\{\varphi,\psi\}_1&=-\left(d_q\varphi,z^{-1}\left[
d_q\psi,L\right]\right),\cr
\{\varphi,\psi\}_2&=\left(q+\Lambda,\left[d_q\varphi,d_q\psi\right]_{R}
\right)-\left(d_q\varphi,\left(d_q\psi\right)^\prime\right),\cr}
\nfr{hamstr}
 where $[x,y]_R\equiv [Rx,y]+[x,Ry]$, with $R(x)=(x_0-x_{<0})/2$, for all
$x\in\hat g$. It can be proved that both brackets satisfy the Jacobi identity
and map  gauge-invariant functionals to a gauge-invariant
functional, see [\Ref{Bd}].
As an aside, the appearance of an R-matrix here is, we
believe, connected with a proper geometric understanding of all of
the hierarchies  above, which  will be reported on elsewhere.
 Also it is found that under time evolution in the coordinate $t_b$,
the following recursion relation holds
 $$
{\partial\varphi\over\partial t_b}=
\{\varphi,H_{zb}\}_1=\{\varphi,H_b\}_2.
\nfr{Av}
We can further show that the Poisson brackets defined
above
sometimes admit non-trivial centres. The existence of these
centres is directly related to the
Hamiltonian densities $h^j$ with $i>j\geq0$. As we have
already remarked these densities
are constant under all the flows of the hierarchy, and so not all the
functionals on ${\cal M}$ are dynamical.

It is well-known that the KdV hierarchies exhibit a scale-invariance, i.e.
under $x\rightarrow \lambda x$, for constant $\lambda$, each quantity
can be assigned scaling dimensions such that the equations remains invariant.
This actually generalizes to all the hierarchies considered above. In fact we
can further generalize this result to arbitrary
transformations $x\rightarrow y(x)$.
This transformation  turns out to be a Poisson mapping of the second symplectic
structure. All this leads us to surmise that the second Poisson bracket
contains (as a subalgebra) the algebra of conformal transformations,
{\it i.e.\/} the Virasoro algebra, which would imply that the second
Poisson bracket algebra is an extended chiral conformal algebra --- thereby
generalizing the occurrence of the $W$-algebras as the second Poisson
bracket algebra of the hierarchies of Drinfel'd and Sokolov.

As an example of the above, consider the first fractional hierarchy
associated to $sl(3)\ $. This corresponds to choosing
$\Lambda=\Lambda_{2,0}$ in \CB. Following our construction, $q$ has
the general form
$$
q=\pmatrix{\star&\star&0\cr \star&\star&\star\cr \star&\star&\star\cr}
+z\pmatrix{0&0&0\cr 0&0&0\cr \star&0&0\cr},
\efr
The gauge invariant functionals are conveniently generated by the
coordinates on the following gauge slice
$$
\tilde q=\pmatrix{\frac12 U&0&0\cr G^+&-U&0\cr T&G^-&\frac12 U\cr}
+\phi\pmatrix{0&1&0\cr 0&0&1\cr z&0&0\cr}.
\efr
The second Poisson bracket algebra is easy to find using \hamstr. One
sees that $\phi$ is in the centre of the algebra, and so may be
consistently set to zero, leaving the $W_3^{(2)}$  algebra
discussed in [\Ref{Sg},\Ref{Se}].
More generally by considering $sl(k)$ with $\Lambda=\Lambda_{j,m}$,
leads to the $W_k^{(j+mk)}$-algebras.

As a final point, we
mention that `Toda' equations can be associated to these hierarchies,
generalizing the relationship between the sine-Gordon equation and the
usual mKdV equation ([\Ref{Si},\Ref{Sc}]) . This will be discussed
elsewhere.

\acknowledgements

The research of JLM was supported by a
Fullbright/MEC fellowship, that of MFdeG by the Natural
Sciences and Engineering Research Council of Canada, that of TJH
by NSF PHY 90-21984 and that of NJB by NSF PHY 86-20266.

\references

\beginref
\Rref{Sa}{S. Novikov, S.V. Manakov, L.P. Pitaevskii and V.E. Zakharov,
`{\sl Theory of
Solitons\/}',
New York, Consultants Bureau 1984\newline
L.D. Faddeev and L.A. Takhtajan, `{\sl Hamiltonian Methods in the Theory of
Solitons\/}',
Springer Series in Soviet Mathematics, Springer-Verlag 1987\newline
A. Das, `{\sl Integrable Models.Singapore\/}', World Scientific 1989;}

\Rref{Sb}{V.G. Kac, `{\sl Infinite Dimensional Lie Algebras\/}', $2^{nd}$
edition. Cambridge University Press 1985;}

\Rref{Sc}{V.G. Drinfel'd and V.V. Sokolov, J. Sov. Math. {\bf 30} (1985) 1975,
Sov. Math. Dokl. {\bf23} (1981) 457;}

\Rref{Sd}{L.J. Mason and G.A.J. Sparling, Phys. Lett. {\bf A137} (1989) 29;}

\Rref{Se}{I. Bakas and D.A. Depireux, Mod. Phys. Lett. {\bf A6} (1991) 1561
(Err
   atum {\bf
A6} (1991) 2351), University of Maryland preprint, UMD-PP91-111(1990);}

\Rref{genws}{F.A. Bias, T. Tjin and P. Van Driel,
Nucl. Phys. {\bf B357} (1991) 632\newline
T. Tjin and P. Van Driel, Amsterdam preprint (1991);}

\Rref{Sf}{A. Newell, `{\sl Solitons in Mathematics and Physics\/}',
 Philadelphia.
Society for Industrial and Applied Mathematics 1985;}

\Rref{Sg}{M. Bershadsky, Comm. Math. Phys. {\bf139} (1991) 71,\newline
A. Polyakov., Int. J. Mod. Phys. {\bf A5} (1990) 833,\newline
P. Mathieu and W. Oevel, Laval University Preprint (1991);}

\Rref{Sh}{E. Brezin and V. Kazakov, Phys. Lett.{\bf
B236}(1990)144\newline M.R. Douglas and S.H. Shenker,
Nucl. Phys. {\bf B 355} (1990) 635\newline D.J. Gross and A.A. Migdal,
Phys. Rev. Lett. {\bf 64} (1990) 127\newline M.R. Douglas, Phys. Lett. {\bf B
238} (1990) 176;}

\Rref{Si}{G.W. Wilson, Ergod. Th. and Dynam. Sys. {\bf
1} (1981) 361;}

\Rref{Sj}{G.W. Wilson, Phil. Trans. R. Soc. Lond. {\bf A315} (1985) 393;}

\Rref{Sk}{R.G. Myhill, `{\sl Automorphisms and Twisted Vertex
Operators\/}', Ph.D. Thesis, Durham 1987;}

\Rref{Sl}{R.S. Ward,
Phil. Trans. R. Soc. Lond. {\bf A 315} (1985) 117\newline M.F. Atiyah,
unpublished\newline N.J. Hitchin,
Proc. Lond. Math. Soc. {\bf 55} (1987) 59;}

\Rref{Sm}{V.G. Kac and D.H. Peterson, in: Symposium on Anomalies, Geometry and
 Topology,
W.A. Bardeen and A.R. White (ed.s). Singapore, World Scientific 1985.}

\Rref{Sn}{J. Lacki,
IAS preprint, IASSNS-HEP-91/6\hfil\newline C. Vafa,
Mod. Phys. Lett. {\bf A 6} (1991)337\newline W. Lerche and N.P. Warner,
Nucl.
Phys. {\bf B358} (1991) 571;}

\Rref{So}{A.M. Polyakov, Mod.Phys.Lett. {\bf A2} (1987) 893\newline
V.G. Knizhnik, A.M. Polyakov and
A.B. Zamolodchikov, Mod. Phys. Lett. {\bf A3} (1988) 819\newline
A.M. Polyakov, Int. J. Mod. Phys. {\bf A5} (1990) 833;}

\Rref{Sp}{Y. Kazama and H. Suzuki,
Phys. Lett. {\bf216B} (1989) 112, Nucl. Phys.
{\bf B321} (1989) 232;}

\Rref{Sq}{M. Douglas, Phys. Lett. {\bf B238 } (1990) 176;}

\Rref{Ra}{T.J. Hollowood and R.G. Myhill, Int. J. Mod. Phys. {\bf A3}
 (1988) 899;}

\Rref{Rb}{V.G. Kac, Functional Analysis and Its Applications {\bf3} (1969)
252;}

\Rref{Rc}{R.W. Carter, Composito
Mathematica. Vol. 25, Fac. 1. (1972) 1;}

\Rref{Rd}{A.P. Fordy, J. Phys. {\bf A17} (1984)
1235;}

\Rref{Re}{V.G. Kac and M. Wakimoto, in: Proceedings of Symposia in Pure
Mathemat
   ics. Vol 49 (1989)
191;}

\Rref{Rf}{V.E. Zakharov and A.B. Shabat, Funkts. Anal. Pril. {\bf8} (1974) 54,
Funkts. Anal. Pril. {\bf13} (1979) 13;}

\Rref{Rg}{N.J. Burroughs, M.F. de Groot, T.J. Hollowood and J.L.
Miramontes,
IAS/Princeton preprint IASSNS-HEP-91/19, PUPT-1251,
to be published in Comm. Math. Phys.;}

\Rref{Rh}{J.F. Adams, `{\sl Lectures on Lie Groups\/}',
Benjamin, New York, 1969;}

\Rref{Ri}{I.M. Gel'fand and L.A. Dikii,
Russ. Math. Surv. {\bf30} (5) (1975) 77, Funkts. Anal. Pril. {\bf 10} (1976)
13,
Funkt. Anal. Pril. {\bf13} (1979) 13;}

\Rref{Rj}{B.A. Kupershmidt and G. Wilson, Invent. Math. 62 (1981) 403-436;}

\Rref{Rk}{O. Babelon and C.M. Viallet, Preprint SISSA-54/89/EP May
1989;}

\Rref{Rl}{V.A. Fateev and A.B. Zamolodchikov, Nucl. Phys. {\bf B280}[FS18]
 (1987)
644;}

\Rref{Rkirillov}{A.A. Kirillov, `{\sl Elements of the Theory of
Representations.\/}', Springer-Verlag 1976;}

\Rref{Rshansky}{M. Semenov-Tian-Shansky, Publ. RIMS. {\bf 21} (1985) 1237;}

\Rref{Bd}{N.J. Burroughs, M.F. de Groot, T.J. Hollowood and J.L.
Miramontes,
IAS and Princeton preprint PUPT-1263, IASSNS-HEP-91/42;}

\endref
\ciao
